\documentclass[graybox]{svmult}

\usepackage{mathptmx}       
\usepackage{amsmath}    
\newcommand{\imu}{\text{i}} 

\usepackage{helvet}         
\usepackage{courier}        
\usepackage{type1cm}        
\usepackage{makeidx}         
\usepackage{graphicx}        
\usepackage{multicol}        
\usepackage[bottom]{footmisc}
\usepackage{epstopdf}

\makeindex             

\begin{document}

\title*{Light-Matter Interactions: \\ A Coupled Oscillator Description}
\author{Martin Frimmer and Lukas Novotny}
\institute{Photonics Laboratory, ETH Z\"urich, 8093 Z\"urich, Switzerland. }

%
\maketitle

\abstract{The semiclassical theory of light-matter interactions describes the interaction between a classical electromagnetic field with a quantum mechanical two-level system. 
We show that the quantum mechanical two-level system can be modeled by a system of two coupled classical harmonic oscillators whose eigenstates are split in frequency according to the coupling strength and play the roles of the two levels of the quantum mechanical two-level system.
The effect of the light field on the mechanical system is modeled as a modulation of the spring constants of the individual oscillators. Using this fully classical model, we derive the Bloch equations for a two-level system and discuss the mechanical analogues of Rabi oscillations and coherent control experiments.}

\section{Introduction}
\label{sec:1}
One of the main thrusts of contemporary physics is quantum engineering, aiming to exploit the properties of quantum systems for information storage, processing and transmission. The fundamental building block of any quantum device is the quantum mechanical two-level system (TLS). In practice, atoms, ions, molecules and solid-state defect centers have been identified as near ideal representations of such a TLS. With the energy-level splittings in these systems corresponding to optical frequencies, light fields provide a handle to control the internal dynamics of such a TLS. To achieve maximum fidelity of the operations on the quantum system, the interaction strength between the electromagnetic field and the TLS has to be maximized by maximizing the field strength at the position of the TLS. Nanophotonics has developed a powerful toolbox to control light at the subwavelength scale, allowing the confinement of electromagnetic radiation to volumes smaller than the limit imposed by diffraction.
With the quality factors of nanophotonic resonators increasing, accompanied by shrinking mode volumes, the interaction strength between a single quantum emitter and a nanophotonic resonator is reaching a level where coherent quantum mechanical effects are observable. Various coherent control schemes, such as Rabi oscillations~\cite{Rabi1937}, Hahn echoes~\cite{Hahn1950}, and Ramsey fringes~\cite{Ramsey1950}, have been developed in the past, based on a semiclassical description of the interaction between a TLS and an electromagnetic field. In this model, the TLS is described quantum mechanically, whereas the light field is of purely classical nature. While this semiclassical theory has been extremely successful in describing the physical reality, it does not provide an intuitive handle to understand the evolution of the quantum mechanical TLS. Interestingly, over the past decades, classical analogues have been constructed for several quantum phenomena~\cite{Dragoman2004}, such as strongly driven two-level systems~\cite{Spreeuw1990}, electromagnetically induced transparency~\cite{Alzar2002}, rapid adiabatic passage~\cite{Shore2009}, and Landau-Zener transitions~\cite{Novotny2010}.
However, a classical Newtonian model describing the internal dynamics of a quantum system driven by an external field has been missing to date.\\

Here, we present a classical model for the interaction of a quantum mechanical TLS with a classical optical field. We construct a \emph{mechanical atom}, consisting of a pair of coupled classical harmonic oscillators. The coupling gives rise to two eigenmodes, split in frequency according to the strength of the coupling between the bare oscillators. These eigenmodes play the role the the two states of a quantum mechanical TLS.
The interaction between the mechanical atom and a driving field is reflected in the modulation of the spring constants of the bare oscillators. Under this parametric driving, the Newtonian equations of motion describing the evolution of our mechanical atom take the exact same form as the optical Bloch equations derived from a semiclassical model based on the Schr\"{o}dinger equation. Accordingly, our model provides as intuitive classical approach to understanding the coherent dynamics of a quantum mechanical TLS.

\section{Semiclassical treatment \label{semiclass}}
\begin{figure}[!b]
\center
\includegraphics[scale=.45]{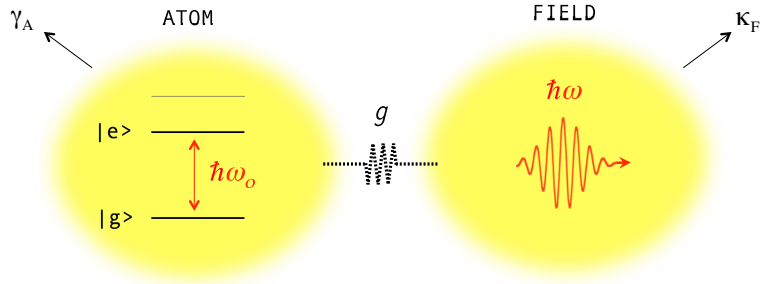}
\caption{Schematic of light-matter interactions. The optical field is characterized by the frequency $\omega$ and the atom (matter) is represented by two electronic states $|g\rangle$ and $|e\rangle$ separated by the energy $\hbar\omega_0$. The interaction of the two systems is characterized by the coupling rate $g$. The excited-state spontaneous decay rate of the atom is $\gamma_A$.
}
\label{fig1}
\end{figure}

Figure~\ref{fig1} conceptually illustrates the interaction between light and matter in a semiclassical framework. The matter part is described by a TLS, from now on termed \emph{atom} for simplicity, with an electronic ground state $|g\rangle$ and an excited state $|e\rangle$.
The two atomic states are separated by the energy $\hbar\omega_0$, with $\omega_0$ the transition frequency. The spontaneous decay rate of the atom is $\gamma_A$. The interaction of the two systems is characterized by the coupling rate $g$, which derives from the interaction Hamiltonian as $g=H_{\rm int}/\hbar$. In the dipole approximation the interaction Hamiltonian can be written as $H_{\rm int}=-{\bf p}\cdot{\bf E}$, with ${\bf p}$ denoting the transition dipole between $|g\rangle$ and $|e\rangle$. Furthermore, ${\bf E}(t)=-{\bf E}_0 \cos \omega t$ is a classical time-harmonic electric field.
Under the influence of the optical field, the wave function of the atom can be written as a superposition of its ground and excited state
\begin{equation}
|\psi\rangle = a(t) |g\rangle + b(t) |e\rangle\; ,
\label{anst}
\end{equation}
where $a(t)$ and $b(t)$ are complex time dependent coefficients. They are found by inserting Eq.~\eqref{anst} into  the time dependent Schr\"{o}dinger equation $\imu\hbar\:\!\partial_t |\psi\rangle=\hat{H} |\psi\rangle$. The detailed procedure is described in textbooks on quantum optics (see, for example, Ref.~\cite{Allen1987}) and we only outline the main aspects here. It is convenient to offset the energy scale, such that the energies of ground state and excited state are $E_g = -\hbar\omega_0/2$ and $E_e = +\hbar\omega_0/2$, respectively, and then move to the rotating frame, that is, performing the  transformation
\begin{equation}
\begin{aligned}
\label{eq:RotFrameAnsatz}
a(t)&\;=\;\bar{a}(t)\exp{[-\imu\omega \,t/2]}\\
b(t)&\;=\;\bar{b}(t)\exp{[+\imu \omega \,t /2]}.
\end{aligned}
\end{equation}
Inserting Eqs.~(\ref{anst}) and (\ref{eq:RotFrameAnsatz}) into the Schr\"odinger equation and performing the rotating wave approximation (i.e. assuming $\omega \sim \omega_0$), we obtain
\begin{equation}
\label{eq:SVEAsystemRotFrame}
\imu\hbar \begin{bmatrix} \dot{\bar{a}} \\ \dot{\bar{b}}\end{bmatrix} \;=\; \frac{\hbar}{2}\begin{bmatrix} \,\;\delta \,& \;\;\;g\,\\  g \,& \!-\delta\end{bmatrix} \begin{bmatrix} \bar{a}\\\bar{b}\end{bmatrix},
\end{equation}
where we have defined the detuning $\delta$ between the driving frequency and the transition frequency
\begin{equation}
\label{eq:detuning}
\delta=\omega_0-\omega\, ,
\end{equation}
and the coupling rate
\begin{equation}
\label{coupl}
g={\bf p}\cdot{\bf E}_0\,/\,\hbar\, ,
\end{equation}
which is also denoted as the classical Rabi frequency. Note that the spontaneous decay rate of the atom does not appear in the semiclassical framework and has to be inserted by hand into the equations of motion in~\eqref{eq:SVEAsystemRotFrame}. A quantized description of the electric field is necessary to retrieve spontaneous emission in a formal fashion~\cite{Haroche2006}. For our purposes, we neglect spontaneous decay, which places our discussion into the regime of strong driving, where $\Omega_R>\gamma_A$ holds for any finite driving field.\\

Using arbitrary initial conditions $\bar{a}(t=0)=\bar{a}_0$ and $\bar{b}(t=0)=\bar{b}_0$, the solutions of Eq.~\eqref{eq:SVEAsystemRotFrame} turn out to be~\cite{Scully1997}
\begin{equation}
\label{eq:solution_ab}
\begin{aligned}
\bar{a}(t)&=\left[-\frac{\imu g}{\Omega_R}\sin\left(\frac{\Omega_Rt}{2}\right)\,\bar{b}_0 + \left\{\cos\left(\frac{\Omega_Rt}{2}\right)-\imu\frac{\delta}{\Omega_R}\sin\left(\frac{\Omega_Rt}{2}\right)\right\}\bar{a}_0\right],\\
\bar{b}(t)&=\left[-\frac{\imu g}{\Omega_R}\sin\left(\frac{\Omega_Rt}{2}\right)\,\bar{a}_0 + \left\{\cos\left(\frac{\Omega_Rt}{2}\right)+\imu\frac{\delta}{\Omega_R}\sin\left(\frac{\Omega_Rt}{2}\right)\right\}\bar{b}_0\right],
\end{aligned}
\end{equation}
where we have introduced the generalized Rabi-frequency
\begin{equation}\label{eq:RabiFreq}
\Omega_R=\sqrt{g^2+\delta^2}.
\end{equation}
Equations~\eqref{eq:solution_ab} together with~\eqref{eq:RotFrameAnsatz} and \eqref{anst} are the general solutions to the problem of a two-level atom interacting with a time-harmonic optical field
${\bf E}(t)=-{\bf E}_0 \cos \omega t$. Before discussing the properties of the solutions we first turn to our purely classical model of the mechanical atom.

\begin{figure}[!b]
\center
\includegraphics[scale=.50]{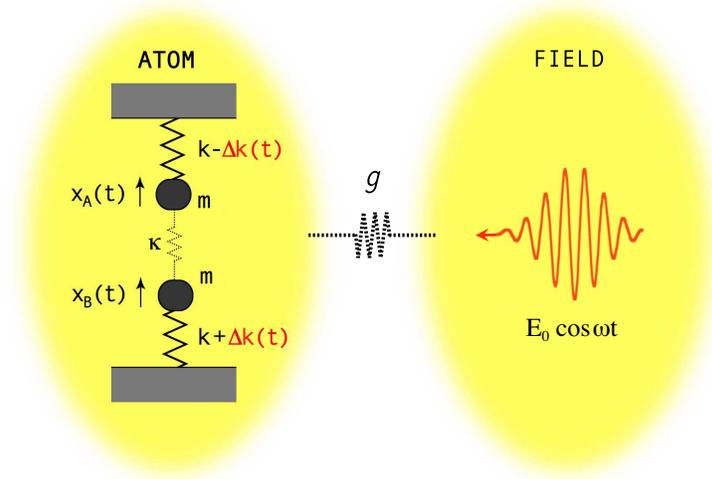}
\caption{Light interacting with a mechanical atom. The mechanical atom consists of two coupled mechanical oscillators with equal masses $m$ and spring constants $k$. The two masses are coupled by a spring with stiffness $\kappa$. The interaction with the optical field gives rise to a parametric modulation of the spring constants $\Delta k$.
}
\label{fig2}
\end{figure}
\section{Classical treatment \label{class}}
We now replace the quantum mechanical atom with states $|g\rangle$ and $|e\rangle$ by a classical mechanical atom made of two coupled classical harmonic oscillators. As shown in Fig.~\ref{fig2}, each oscillator consists of a mass $m$ suspended by a spring with spring constant $k$ and the oscillators are coupled by a spring with stiffness $\kappa$. The interaction with the optical field is described by a parametric process, that is, the optical field modulates the spring constant $k$ of one of the oscillators by an amount $\Delta k(t)$ and that of the other by $-\Delta k(t)$. The solution of this coupled system has been recently derived in Ref.~\cite{Frimmer2014}. Here we outline the main steps and show that the solutions of this system are identical to those found for the semiclassical treatment in the previous section.\\

In terms of the coordinates $x_A$ and $x_B$ of the two oscillators, the equations of motion are
\begin{equation}
\label{eq:coupledosc}
\begin{aligned}
m \;\!\ddot{x}_A + [k+\kappa - \Delta k(t)] \;\!x_A - \kappa \;\!x_B &= 0\, ,  \\
m \;\!\ddot{x}_B + [k+\kappa + \Delta k(t)] \;\!x_B - \kappa \;\!x_A &= 0\, ,
\end{aligned}
\end{equation}
where $m$ are the masses, $k$ the spring constants, $\Delta k(t)$ a time-dependent modulation of the spring constants, and $\kappa$ the coupling spring constant. We introduce the carrier frequency $\Omega_0$, the detuning frequency $\Omega_d$ and the coupling frequency $\Omega_c$ as
\begin{equation}
\label{eq:frequencies}
\begin{aligned}
\Omega_0^2&= [k+\kappa] / m,\\
\Omega_d^2&= \Delta k / m,\\
\Omega_c^2&= \kappa/m,
\end{aligned}
 \end{equation}
 and represent the coupled differential equations in~\eqref{eq:coupledosc} in matrix form as
\begin{equation}
\label{eq:coupledoscMatrixForm}
\left[\frac{d^2}{dt^2} + \Omega_0^2\right]
\begin{bmatrix} x_A \\ x_B \end{bmatrix} + \begin{bmatrix} -\Omega_d^2 & - \Omega_c^2\\ -\Omega_c^2 & \;\,\Omega_d^2\end{bmatrix}\begin{bmatrix} x_A \\ x_B \end{bmatrix}
\;=\; 0\; .
\end{equation}
This system of equations describes the full dynamics of the undriven coupled oscillator problem. \\

\subsubsection*{Eigenmodes of the coupled oscillator}
We now solve for the eigenmodes of the system and their respective eigenfrequencies in absence of modulation ($\Omega_d = 0$). To diagonalize the matrix in Eq.~\eqref{eq:coupledoscMatrixForm} we write
\begin{equation}
\label{eq:transformEigenmodes}
\begin{bmatrix} x_A\\x_B\end{bmatrix} = \begin{bmatrix} 1 & \;\;1 \\
1 & -1 \end{bmatrix} \begin{bmatrix} x_{+} \\ x_{-} \end{bmatrix},
\end{equation}
which yields two independent differential equations for the normal mode coordinates $x_+$ and $x_-$
\begin{equation}
\left[\frac{d^2}{dt^2} + \Omega_+^2\right] x_{+} = 0\, \quad \left[\frac{d^2}{dt^2} + \Omega_-^2\right] x_{-} = 0\, , 
\label{eq:decoupledHarmOsc}
\end{equation}
with the eigenfrequencies
\begin{equation}
\Omega_{\pm} =  \left[\Omega_0^2 \mp \Omega_c^2\,\right]^{1/2}.
\end{equation}
Here, $\Omega_+$ denotes the frequency of the symmetric eigenmode ($x_+ = x_A + x_B$), which is lower than the frequency $\Omega_-$ of the antisymmetric eigenmode ($x_- = x_A - x_B$).  The frequency splitting is
\begin{equation}
\omega_0\;=\; \Omega_--\Omega_+ \approx \frac{\Omega_c^2}{\Omega_0},
\end{equation}
where we made use of $\Omega_c\ll\Omega_0$. Thus, the splitting is proportional to the coupling strength $\kappa$. \\

\subsubsection*{Interaction with the optical field}
We now transform the equations of motion \eqref{eq:coupledoscMatrixForm} to the basis $x_+, x_-$ and obtain
\begin{equation}
\label{eq:coupledoscMatrixFormEigen}
\left[\frac{d^2}{dt^2} + \Omega_0^2\right]
\begin{bmatrix} x_+ \\ x_- \end{bmatrix} + \begin{bmatrix} -\Omega_c^2 & - \Omega_d^2\\ -\Omega_d^2 & \;\,\Omega_c^2\end{bmatrix}\begin{bmatrix} x_+ \\ x_+ \end{bmatrix}
\;=\; 0 \; .
\end{equation}
We demand that the interaction of the mechanical atom with the optical field oscillating at frequency $\omega$ gives rise to a modulation of the oscillators' spring constants
\begin{equation}
\label{eq:parametricDriving}
\Delta k(t)= 2 m \Omega_0  g\cos(\omega t),
\end{equation}
where $g$ is the coupling rate. To understand the evolution of the eigenmodes we write
\begin{equation}
\label{eq:SVEAansatz}
\begin{aligned}
x_+ & =\text{Re}\left\{a(t)\exp{[\imu\Omega_0t]}\right\},\\
x_- & =\text{Re}\left\{b(t)\exp{[\imu\Omega_0t]}\right\},
\end{aligned}
\end{equation}
where each mode is rapidly oscillating at the carrier frequency $\Omega_0$ and modulated by the slowly varying complex amplitudes $a(t)$ and $b(t)$, respectively. Upon inserting~\eqref{eq:SVEAansatz} into the coupled equations of motion~\eqref{eq:coupledoscMatrixFormEigen} we assume that the amplitude functions $a(t)$ and $b(t)$ do not change appreciably during an oscillation period $2\pi/\Omega_0$, which allows us to neglect terms containing second time derivatives (slowly varying envelope approximation). With this approximation we arrive at the following equations of motion for the eigenmode amplitudes
\begin{equation}\label{eq:SVEAsystem}
\imu\begin{bmatrix} \dot{a} \\ \dot{b}\end{bmatrix} = \frac{1}{2}\begin{bmatrix}\omega_0  &-\Omega_d^2/\Omega_0\\  -\Omega_d^2/\Omega_0 & -\omega_0 \end{bmatrix} \begin{bmatrix} a\\b\end{bmatrix}\, .
\end{equation}
In a next step we apply the transformation
\begin{equation}
\begin{aligned}
\label{eq:RotFrameAnsatz2}
a(t)&\;=\;\bar{a}(t)\exp{[-\imu\omega \,t/2]}\\
b(t)&\;=\;\bar{b}(t)\exp{[+\imu \omega \,t /2]}.
\end{aligned}
\end{equation}
Here, $\bar{a}$ and $\bar{b}$ are the slowly varying amplitudes of the symmetric and antisymmetric eigenmodes  in a coordinate frame rotating at the driving frequency.
This transformation generates terms $\exp[\pm3\imu\omega t/2]$ in~\eqref{eq:SVEAsystem}, which are rapidly oscillating and which we neglect since they average out on the time scales of interest (rotating wave approximation). In terms of the detuning $\delta=\Delta\Omega-\omega$ we then obtain
\begin{equation}
\label{eq:SVEAsystemRotFrame2}
\imu \begin{bmatrix} \dot{\bar{a}} \\ \dot{\bar{b}}\end{bmatrix} \;=\; \frac{1}{2}\begin{bmatrix} \,\;\delta \,& \;\;\;g\,\\  g \,& \!-\delta\end{bmatrix} \begin{bmatrix} \bar{a}\\\bar{b}\end{bmatrix},
\end{equation}
which is identical with~(\ref{eq:SVEAsystemRotFrame}) derived in the previous section using semiclassical theory. Hence, the solutions of~(\ref{eq:SVEAsystemRotFrame2}) are given by~(\ref{eq:solution_ab}). We conclude that the semiclassical theory of light-matter interaction can be reproduced by a pair of parametrically modulated classical oscillators. Note that both theories make use of the rotating wave approximation, which is reflected by the fact that~\eqref{eq:solution_ab} only retains dynamics on the time scale given by the generalized Rabi-frequency $\Omega_R$ and neglects any fast dynamics on time scales set by the optical frequency $\omega$. Accordingly, our solutions are only valid for coupling rates $g$ and detunings $\delta$ small enough to ensure $\Omega_R\ll\omega$. \\

\section{Arbitrary interaction phases}
In section~\ref{semiclass} we have considered an optical field of the form ${\bf E}(t)=-{\bf E}_0 \cos(\omega t)$. We now allow for an arbitrary phase offset $\phi_0$, such that ${\bf E}(t)=-{\bf E}_0 \cos(\omega t-\phi_0)=-{\bf E}_0 \cos\phi_0 \cos(\omega t) - {\bf E}_0 \sin\phi_0\sin(\omega t)$. This turns the interaction Hamiltonian into
\begin{equation}
H_{\rm int} \,=\, -{\bf p}\cdot{\bf E} \; = \; \hbar  \left[g_x \cos(\omega t)\,+\, g_y\,\sin(\omega t)\right] \, ,
\end{equation}
where we defined the coupling rates as $g_x = g \cos\phi_0$ and $g_y = g \sin\phi_0$, with $g={\bf p}\cdot{\bf E}_0/\hbar$. Correspondingly, in the classical oscillator model considered in section~\ref{class}, the modulation of the spring constant becomes
\begin{equation}
\Delta k \;=\; 2 m \Omega_0 \left[g_x \cos(\omega t)\,+\, g_y\,\sin(\omega t)\right] \, .
\end{equation}
As a result of the phase $\phi_0$, the systems of equations~(\ref{eq:SVEAsystemRotFrame}) and~(\ref{eq:SVEAsystemRotFrame2}) turn into
\begin{equation}
\label{eq:SVEAsystemRotFrame5}
\imu\hbar \begin{bmatrix} \dot{\bar{a}} \\ \dot{\bar{b}}\end{bmatrix} \;=\; \frac{\hbar}{2}\begin{bmatrix} \,\;\delta \,& \;\;\;g_x-\imu g_y\,\\  g_x+\imu g_y \,& \!-\delta\end{bmatrix} \begin{bmatrix} \bar{a}\\\bar{b}\end{bmatrix}.
\end{equation}

\subsubsection*{Representation in terms of Pauli matrices}
Let us write the eigenstates of the atom as vectors
\begin{equation}
\label{vectrep}
|g\rangle = \begin{bmatrix} 1 \\ 0\end{bmatrix}\, ,  \quad |e\rangle = \begin{bmatrix} 0 \\ 1\end{bmatrix} \; ,
\end{equation}
which turn the state of the atom in the rotating frame into
\begin{equation}
\label{vectrep2}
|\bar{\psi}\rangle = \bar{a}(t) |g\rangle + \bar{b}(t) |e\rangle = \begin{bmatrix} \bar{a} \\ \bar{b}\end{bmatrix}\; .
\end{equation}
The evolution of $|\bar{\psi}\rangle$ is governed by the Schr\"odinger equation $\imu\hbar\:\!\partial_t |\psi\rangle = {\bar{H}}  |\bar{\psi}\rangle$
with $\bar{H}$ being the Hamiltonian in the rotating frame. Inserting~\eqref{vectrep2} yields
\begin{equation}
\label{schrd2}
\imu\hbar  \begin{bmatrix} \dot{\bar{a}} \\ \dot{\bar{b}}\end{bmatrix} \,=\, {\bar{H}}  \begin{bmatrix} \bar{a} \\ \bar{b}\end{bmatrix}\;
\end{equation}
and a comparison with~\eqref{eq:SVEAsystemRotFrame5} gives
\begin{equation}
\label{eq:SVEAsystemRotFrame3}
\bar{H} \;=\; \frac{\hbar}{2}\begin{bmatrix} \,\;\delta \,& \;\;\;g_x-\imu g_y\,\\  g_x+\imu g_y \,& \!-\delta\end{bmatrix} \; .
\end{equation}
In terms of the Pauli matrices
\begin{equation}
\label{pauli}
{\sigma_x} \;=\; \begin{bmatrix} 0 \;&\; 1 \\  1 \;&\;  0\end{bmatrix}\, ,\quad {\sigma_y} \;=\; \begin{bmatrix} 0 \;& \!-\imu \\  \imu \;&\;  \,0\end{bmatrix} \, ,\quad {\sigma_z} \;=\; \begin{bmatrix} 1 \;&\; \,0 \\  0 \;&  \!-1\end{bmatrix} \; ,
\end{equation}
we can cast $\bar{H}$ in the compact form
\begin{equation}
\label{eq:SVEAsystemRotFrame4}
\bar{H} \;=\; \frac{\hbar\:\!g_x}{2} \sigma_x\,+\,\frac{\hbar\:\!g_y}{2} \sigma_y\,+\, \frac{\hbar\:\!\delta}{2} \sigma_z \;=\; \hbar {\bf R}\cdot \mbox{\boldmath{$\sigma$}}/2 \; ,
\end{equation}
where we introduced the rotation vector ${\bf R}=[g_x,g_y,\delta]^T$ and defined $\mbox{\boldmath{$\sigma$}} = [\sigma_x,\sigma_y,\sigma_z]^T$.

\begin{figure}[!b]
\sidecaption
\includegraphics[scale=.60]{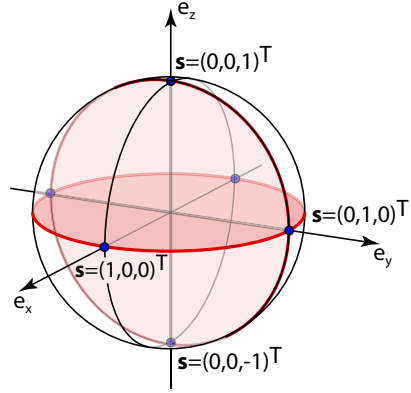}
\caption{The Bloch sphere. A pair of complex amplitudes ($\bar{a},\bar{b}$) is represented by the Bloch vector ${\bf s}=(s_x,s_y,s_z)^T$.  Normalized amplitude pairs  ($|\bar{a}|^2+|\bar{b}|^2=1$) lie on the surface of the Bloch sphere, which has unit radius. All amplitude pairs $(\bar{a},\bar{b})\exp[\imu\varphi]$ with arbitrary $\varphi$ are mapped onto the same point ${\bf s}$.
}
\label{fig3}
\end{figure}
\section{The Bloch equations}
Let us represent the solutions $\bar{a}$ and $\bar{b}$ in a vectorial form that was originally introduced by Felix Bloch in the context of nuclear magnetic resonance~\cite{Bloch1946}. To do so,
we introduce the Bloch vector ${\bf s}=(s_x,s_y,s_z)^T$ with the components
\begin{equation}
\label{eq:blochvectoreq}
\begin{aligned}
s_x &= \hphantom{\imu(}\bar{a} \bar{b}^{\ast}  +  \bar{a}^{\ast}\bar{b}        &=&\hphantom{-}2\,  {\rm Re}\{\bar{a}\bar{b}^{\ast}\}   &=\hphantom{-}2|\bar{a}||\bar{b}|\cos\phi\, ,\\
s_y &= \imu (\bar{a} \bar{b}^{\ast} - \bar{a}^{\ast}\bar{b})                   &=&{-}2\,             {\rm Im}\{\bar{a}\bar{b}^{\ast}\}              &= -2|\bar{a}||\bar{b}|\sin\phi\, ,\\
s_z &= \hphantom{\imu(} \bar{a} \bar{a}^{\ast} -  \bar{b}\bar{b}^{\ast}        &=&\hphantom{-}     |\bar{a}|^2 - |\bar{b}|^2.               &
\end{aligned}
\end{equation}
The Bloch vector ${\bf s}$ encodes in its three real-valued components the state of the atom, which is represented by the amplitudes $|\bar{a}|, |\bar{b}|$ and the relative phase $\phi$. Importantly, every state $(\bar{a},\bar{b})$ of the oscillator system can be multiplied by an arbitrary phase factor $\exp[\imu\varphi]$ without changing the corresponding Bloch vector ${\bf s}$. Discarding this absolute phase of the complex amplitudes $\bar{a},\bar{b}$ reduces the degrees of freedom from four (two real amplitudes and two phases for $\bar{a}$ and $\bar{b}$) to three, such that the state of the oscillator system can be represented in the three dimensional Bloch vector space.\\

For a normalized system ($|\bar{a}|^2+|\bar{b}|^2=1$) the tip of the Bloch vector always lies on a unit sphere, called the Bloch sphere, illustrated in Fig.~\ref{fig3}.
The north pole of the Bloch sphere ${\bf s}=(0,0,1)^T$ corresponds to the state vector $(\bar{a},\bar{b})=(1,0)$,  the ground state $|g\rangle$ of the atom, according to Eq.~\eqref{anst}. Accordingly, for the excited state $|e\rangle$, corresponding to $(\bar{a},\bar{b})=(0,1)$, the tip of the Bloch vector is located at the south pole of the Bloch sphere ${\bf s}=(0,0,-1)^T$. All points on the equator of the Bloch sphere correspond to equal superpositions of $|g\rangle$ and $|e\rangle$, but with varying relative phase $\phi$. For example, the state $(\bar{a},\bar{b})=(1,1)/\sqrt{2}$ lies at the intersection of the $x$-axis and the Bloch sphere ${\bf s}=(1,0,0)^T$, whereas the state $(\bar{a},\bar{b})=(1,\imu)/\sqrt{2}$ lies at the intersection with the $y$-axis ${\bf s}=(0,1,0)^T$.\\

It is instructive to express the dynamics of the light-matter interaction in terms of the Bloch vector ${\bf s}$.
Using Eqs.~\eqref{eq:SVEAsystemRotFrame3} and~\eqref{eq:blochvectoreq} we can easily show that the time evolution of the Bloch vector is given by
\begin{equation}\label{eq:BlochEqMotion}
\frac{d}{d t}\begin{bmatrix} s_x \\ s_y \\ s_z\end{bmatrix} = \begin{bmatrix}  \;\;0 & \!-\delta & \;\;\,g_y \\  \;\;\delta & \;\;0 & -g_x \\ -g_y & \;\;\;g_x & \;\;\,0 \end{bmatrix} \begin{bmatrix} s_x \\ s_y \\ s_z\end{bmatrix} .
\end{equation}
This system of equations can be represented in compact form as
\begin{equation}
\label{eq:BlochEqMotion2}
\dot{\bf s} \;=\; {\bf R} \times {\bf s},
\end{equation}
where ${\bf R}=(g_x, g_y, \delta)^T$. The equation of motion~(\ref{eq:BlochEqMotion2}) describes the precession of the Bloch vector ${\bf s}$ around the rotation vector ${\bf R}$ with the angular frequency  $\Omega_R$ defined in Eq.~\eqref{eq:RabiFreq}, where $\Omega_R$  equals the length of ${\bf R}$. \\

A point on the Bloch sphere entirely defines the state of the atom. According to Eq.~\eqref{eq:BlochEqMotion2} we can bring the atom from any starting point to any other point on the Bloch sphere simply by choosing the right rotation vector ${\bf R}$ and waiting for the right time to achieve the desired amount of rotation. This idea is at the core of the concept of coherent control.
We note that we have neglected spontaneous emission both in our semiclassical treatment and in the mechanical atom. Due to spontaneous emission, a quantum two-level system will always end up in its ground state after a long time. Spontaneous emission is a process that is genuinely quantum mechanical in nature and requires a fully quantized theory for the electromagnetic field. Even Bloch added the decay constants semi-phenomenologically in his treatment of nuclear spins\cite{Allen1987}. \\

In 1937 Rabi studied the dynamics of a spin in a static magnetic field that is modulated by a radio frequency field and he found that the spin vector is periodically oscillating between parallel and anti-parallel directions with respect to the static magnetic field~\cite{Rabi1937}. These oscillations are referred to as Rabi oscillations, or Rabi flopping. To illustrate Rabi oscillations we assume a resonant ($\delta=0$)  driving with $g_y=0$, such that the Bloch vector, starting at the north pole ${\bf s}=(0,0,1)^T$, rotates around the axis ${\bf R}=g\:\!{\bf e}_x$ at a frequency $\Omega_R=g_x\,$ according to Eq.~\eqref{eq:BlochEqMotion2}. After a time $t_\pi=\pi/g$ the Bloch vector will have rotated to the south pole ${\bf s}=(0,0,-1)^T$. Obviously, parametric driving for a time $t_\pi$ (called $\pi$-pulse) inverts our system.  Accordingly, after parametrically driving the system for a time $t_{2\pi}=2\pi/g$ it has returned to its initial state at the north pole of the Bloch sphere. For a continuous parametric driving, starting at ${\bf s}=(0,0,1)^T$, the system is oscillating between its two eigenmodes at the resonant Rabi-frequency $\Omega_R=g_x$. \\

If we start out at the north pole ${\bf s}=(0,0,1)^T$ but use a resonant driving with $g_x=0$, the Bloch vector will rotate around the axis ${\bf R}=g\:\!{\bf e}_y$ at a frequency $\Omega_R=g_y\,$. Thus, by selectively switching on and off the parameters $g_x$, $g_y$ and $\delta$ we can rotate the Bloch vector around any axis and by an arbitrary amount.

\section{Discussion}
We have reviewed the semiclassical theory of light-matter interaction and presented a classical oscillator model that yields identical equations of motion for the amplitudes of a two-level system. The oscillator model comprises two coupled harmonic oscillators with spring constants that are modulated by the external optical field. The coupling strength $\kappa$ between the two oscillators sets the transition frequency $\omega_0$ and defines the detuning $\delta$. The correspondence between the classical and the quantum system is established by using the slowly varying envelope approximation, which casts the Newtonian equations of motion of the coupled oscillators into a form resembling the Schr{\"o}dinger equation for a two-level atom. The mechanical analogue lends itself to visualize optical experiments, such as Rabi oscillations, Ramsey fringes and Hahn echoes. In a continuation of the work presented here it would be interesting to consider a model of {\em strong} light-matter interactions, in which the driving optical field is itself modeled by a mechanical oscillator.
%

\end{document}